\documentstyle[aps,epsfig]{revtex}
\begin{document}
\title{ Modified Bethe-Weizs\"acker mass formula with isotonic shift and new driplines }

\author{P. Roy Chowdhury$^1$, C. Samanta$^{1,2}$ \thanks{E-mail:chhanda@lotus.saha.ernet.in}}
\address{ $^1$ Saha Institute of Nuclear Physics, 1/AF Bidhan Nagar, Kolkata 700 064, India }
\address{ $^2$ Physics Department, Virginia Commonwealth University, Richmond, VA 23284-2000, U.S.A. }

\author { D.N. Basu}
\address {Variable  Energy  Cyclotron  Centre,  1/AF Bidhan Nagar, Kolkata 700 064, India}
\date{\today }
\maketitle
\begin{abstract}

      Nuclear masses are  calculated using the modified Bethe-Weizs\"acker mass formula in which the isotonic shifts have been incorporated. The results are compared with the improved liquid drop model with isotonic shift. Mass excesses predicted by this method compares well with the microscopic-macroscopic model while being much more simple. The neutron and proton drip lines have been predicted using this modified Bethe-Weizs\"acker mass formula with isotonic shifts.

\vskip 0.2cm
\noindent
Keywords : Mass formula, Drip Lines, Separation energies, Exotic nuclei.

\noindent  
PACS numbers:27.30.+t, 21.10.Dr, 32.10.Bi    
\end{abstract}
\vskip 0.2cm

      Nuclear reactions using radioactive ion beams have been studied extensively to allow measurements of the masses, half-lives, radii and other properties of unstable nuclei. These results reveal some interesting characteristics of unstable nuclei, such as a halo and a skin \cite{r1}. The disappearance of some traditional magic numbers and extra stability for some neutron numbers \cite{r2} have also been observed. The  Bethe-Weizs\"acker (BW) mass formula designed to reproduce the gross features of nuclear binding energies for medium and heavy nuclei fails for light nuclei, specially away from the line of stability \cite{r3}. The unusual stability of nuclei with preferred nucleon numbers, commonly referred to as magic numbers, can be clearly delineated by comparing the experimental binding energies with predictions of the Bethe-Weizs\"acker mass formula. However, this comparison does not indicate the recently observed features like the disappearance of some traditional magic numbers and extra stability for some newly observed nuclei. To alleviate these problems, a generalized Bethe-Weizs\"acker mass formula was predicted \cite{r4} which explained the gross features of the shapes of the binding energy versus neutron number curves of all the elements from $Li$ to $Bi$. Both Bethe-Weizs\"acker and the modified  Bethe-Weizs\"acker (BWM) mass formulae have no shell effect incorporated and thus when the shell effect in a nucleus quenches, the theoretical mass comes closer to the experimental ones. A comparison with the experimental data and the BWM delineates the positions of all old and the new magic numbers. For heavier nuclei this BWM mass formula approaches the old BW mass formula. 

      In this work we improve the mass predictions of BWM by introducing the isotonic mass shifts \cite{r5}. We find that it greatly improves the agreement between the experimental masses and the calculated ones and thus predicts the driplines more accurately than calculated earlier \cite{r6} with BWM alone. We adopt a procedure in which we compute the average shift of the BWM formula from the experimental values. The shift, called the isotonic shift $\Delta_n$, is then subtracted from the prediction of the BWM formula \cite{r5}, as explained later.

      In the liquid drop description of a nucleus of mass number A and atomic number Z, the binding energy is given by,

\begin{equation}
 B(A,Z) = a_vA-a_sA^{2/3}-a_cZ(Z-1)/A^{1/3}-a_{sym}(A-2Z)^2/A+\delta,
\label{seqn1}
\end{equation}
\noindent
where

\begin{equation}
 a_v=15.78~MeV,~a_s=18.34~MeV,~a_c=0.71~MeV,~a_{sym}=23.21~MeV~and~a_p=12~MeV,
\label{seqn2}
\end{equation}
\noindent
and the pairing term,

\begin{eqnarray}
  \delta=&&a_pA^{-1/2}~for~even~N-even~Z,\nonumber\\
         =&&-a_pA^{-1/2}~for~ odd~N-odd~Z,\nonumber\\
         =&&0~for~odd~A,\nonumber\\
\label{seqn3}
\end{eqnarray} 
\noindent

    The above formula was prescribed by Bethe-Weizs\"acker (BW). With the discovery of neutron rich light nuclei, it was found that the BW is severely inadequate for light mass nuclei away from the valley of stability. From a phenomenological search a more generalized BW formula was predicted by Samanta and Adhikari \cite{r4} in which the binding energy is given by,   

\begin{equation}
 B(A,Z) = a_vA-a_sA^{2/3}-a_cZ(Z-1)/A^{1/3}-a_{sym}(A-2Z)^2/[(1+e^{-A/k})A]+\delta_{new},
\label{seqn4}
\end{equation}
\noindent
where,

\begin{equation}
 \delta_{new}=(1-e^{-A/c})\delta,~~~~c=30,~~~~k=17,
\label{seqn5}
\end{equation}
\noindent
and the constant $a_v=15.777 MeV$ \cite{r7}. The other constants remain the same and $\delta$ is the old pairing term. 

      We made five parameter searches for the BW mass formula using Audi-Wapstra 2003 \cite{r8} mass table by minimizing $\chi^2/N$ and the root mean square deviation $\sigma$ for atomic mass excesses. The results are listed in Table-1 below. We find that both and $\chi^2/N$ and the root mean square deviation $\sigma$, which are defined later in the text, can not be minimized simultaneously with the same set of parameters. These five parameters obtained from $\chi^2/N$ minimizations are the same as those used in BWM \cite{r7}. Once the five parameters are fixed from the $\chi^2/N$ minimization, the rest two parameters c and k were extracted by matching the experimental binding energy versus neutron number curves from $Li$ to $Bi$. For all the 2518 nuclei available in Audi-Wapstra 2003 \cite{r8} mass table from $Li$ to $Bi$, excluding $^{12}C$ and $^{20,21}F$ for which the experimental mass excesses are either zero or very small, the values obtained are $(\chi^2/N)_{BW}=3.250$, $(\chi^2/N)_{BWM}=1.434$, $(\sigma)_{BW}=4.711 MeV$ and $(\sigma)_{BWM}=3.788 MeV$. 

      As BWM mass fomula does not have shell effect incorporated, it can not predict the masses of the magic nuclei and nuclei with deformities. Also, as there is no Wigner term, it fails to predict the extra stability of the N=Z nuclei. To predict the masses of all the nuclei more accurately, we introduce the isotonic shift \cite{r5} in the mass calculations with BWM. The isotonic shift corrected binding energy, $BE_{cor}(A, Z)$, is defined as,  
                   
\begin{equation}
 BE_{cor}(A, Z) = BE_{Th}(A, Z) - \Delta_n,
\label{seqn6}
\end{equation}
\noindent
where

\begin{equation}
 \Delta_n = (1/n) \sum_i [BE_{Th}(A_i , Z_i) - BE_{expt}(A_i, Z_i) ],
\label{seqn7}
\end{equation}
\noindent
and n=3 is the number of points taken before the right or left end of the curve. The $ BE_{Th}(A, Z)$ is the theoretical binding energy. To achieve the most accurate treatment of the masses, we employ the recently available masses given in the Audi-Wapstra mass table, 2003 \cite{r8}. There still exists a large number of unknown masses for some nuclei, most of which have extreme proton to neutron ratios. For example $^{61}As$ is the heaviest and  $^{40}Mg$ is the lightest isotones for N=28 listed in the Audi-Wapstra table 2003 \cite{r8}. To predict the binding energy for N=28 isotone heavier than $^{61}As$, viz. $^{62}Se$, we use the masses of $^{59}Ga(Z=31)$, $^{60}Ge(Z=32)$ and $^{61}As(Z=33)$ and eqns. (6) and (7). Similarly, we compute $\Delta_n$  from the masses of $^{40}Mg(Z=12)$, $^{41}Al(Z=13)$, $^{42}Si(Z=14)$ to predict the mass of $^{39}Na$ which is a N=28 isotone lighter than $^{40}Mg$. To accomplish this computationally we find that the formula 
\begin{equation}
 N_s=0.968051 Z+0.00658803 Z^2,
\label{seqn8}
\end{equation}
\noindent
provides a reasonable description of the neutron number $N_s$ of a nucleus with atomic number Z on the line of beta-stability. For a particular (Z,N) nucleus, $N_s$ decides which side of the line of beta stability  the chosen nucleus lies. 

\begin{table}
\caption{The parameters for the Bethe-Weizs\"acker mass formula found by fitting the available atomic mass  excesses of Audi-Wapstra 2003 [8] mass table.}
\begin{tabular}{cccccc}
$a_v$ & $a_s$ & $a_c$ & $a_{sym}$ & $a_p$ & Minimized   \\
$ MeV$ & $MeV$ & $ MeV$ & $MeV$ & $ MeV$ & quantity  \\ \hline
&&&&&\\
15.777(53)&18.34(17)&0 .710(3)&23.21(10)&12(2)&$\chi^2/N$=3.34\\
&&&&&\\
15.258(20)&16.26(6)& .689(1)&22.20(5)&10.08(85)&$\sigma^2$=15.12 \\
&&&&&\\
\end{tabular} 
\end{table}
\nopagebreak

      The one proton and two proton separation energies $S_{1p}$ and $S_{2p}$ defined as the energies required to remove one proton and two protons from a nucleus are given by,

\begin{equation}
 S_{1p} = B(A,Z) - B(A-1,Z-1)~~~~and~~~~S_{2p} = B(A,Z) - B(A-2,Z-2)
\label{seqn9}
\end{equation}   
\noindent
respectively, while the one neutron and two neutron separation energies $S_{1n}$ and $S_{2n}$ defined as the energies required to remove one neutron and two neutrons from a nucleus are given by,

\begin{equation}
 S_{1n} = B(A,Z) - B(A-1,Z)~~~~and~~~~S_{2n} = B(A,Z) - B(A-2,Z)
\label{seqn10}
\end{equation}   
\noindent
respectively. 

      The atomic mass excess $\Delta M_{A,Z}$ for  a nucleus with mass number A and atomic number Z is given by 

\begin{equation}
 \Delta M_{A,Z} =  Z \Delta m_H  + (A-Z) \Delta m_n - a_{el} Z^{2.39} - B(A,Z)
\label{seqn11}
\end{equation}
\noindent
where $\Delta m_H = m_p + m_e - u$ =  7.28898483 MeV and $\Delta m_n = m_n - u$ = 8.07131710 MeV, $m_p$, $m_n$ and $m_e$ are the masses of proton, neutron and electron respectively, the atomic mass unit u is 1/12 the mass of $^{12}C$ atom, $\Delta M_{A,Z}$ is the atomic mass excess of an atom of mass number A and atomic number Z and the electronic binding energy constant $a_{el} = 1.433 \times 10^{-5}$ MeV.

       In the present work the one neutron and one proton separation energies have been calculated using the newly extended Bethe-Weizsacker (BWM) mass formula given by eqn. (4) with the isotonic shift, called BWMIS. The last stable nucleus from which the removal of a single neutron (and any more) makes the one proton separation energy negative defines a proton drip line nucleus and the last stable nucleus to which addition of a single neutron (and any more) makes the one neutron separation energy negative defines a neutron drip line nucleus. Proton and neutron drip line nuclei, thus obtained, over the entire (N,Z) region define the predicted neutron and proton drip lines.

      The two quantities of importance that can be used for a relative comparison between different theoretical models are $\chi^2/N$ and the root mean square deviation $\sigma$ which are defined as 

\begin{equation}
 \chi^2/N =  (1/N) \sum [ ( \Delta M_{Th} - \Delta M_{Ex} ) / \Delta M_{Ex} ]^2
\label{seqn12}
\end{equation}  
\noindent
and

\begin{equation}
 \sigma^2 =  (1/N) \sum [ \Delta M_{Th} - \Delta M_{Ex} ]^2
\label{seqn13}
\end{equation}   
\noindent
respectively, where $\Delta M_{Th}$ is the theoretically calculated atomic mass excess and the summation extends to $N$ data points for which experimental atomic mass excesses are known. For the calculations of $\chi^2/N$ and the root mean square deviation $\sigma$, experimental masses \cite{r8} of 3175 nuclei have been used for the cases of improved liquid drop model (ILDM) \cite{r5} and BWM. For the case of Myers-Swiatecki (MS) \cite{r9}, out of 3175 experimental masses, theoretical masses of 23 nuclei are not available. Therefore, we use only experimental masses of 3152 nuclei for calculating the $\chi^2/N$ and $\sigma$ for MS masses. The values $\chi^2/N$ obtained for the MS, BWM and ILDM are 1.398, 8.203 and 20.525 respectively, while $\sigma$ for those are 0.865 MeV, 4.247 MeV and 3.061 MeV respectively. When isotonic shifts have been incorporated in the BWM and ILDM mass formulae, the $\chi^2/N$ reduces to 7.758 and 10.950 respectively while $\sigma$ becomes 1.645 MeV and 1.529 MeV for BWM and ILDM respectively. In case of BWM the isotonic shifts have been incorporated beyond A=30 since the BWM \cite{r4} was designed to improve the predictions of low mass region. It is worthwhile to mention here that $\sigma$ is a measure of absolute error while the $\chi^2/N$ is a measure of relative error. In many cases, such as reaction Q value or separation energy and dripline calculations,  absolute errors involved in atomic mass excesses are the  quantities of concern and $\sigma$ plays more important role than $\chi^2/N$. On the contrary, $\chi^2/N$ signifies the relative, that is, uniform percentage error involved in the mass predictions. In fact, in BWMIS the value of $\chi^2/N$ reduces from 7.758 to 0.907 if just 2 nuclei, namely $^{20}F$ and $^{21}F$, for which experimental mass excesses are of the order of $10^{-2} MeV$, are excluded from the calculations. This happens as in the computation of the $\chi^2/N$, the experimental mass excesses $\Delta M_{Ex}$ appear in the denominator and any small deviation of $\Delta M_{Th}$ from $\Delta M_{Ex}$ is blown up excessively when divided by a number close to zero. But it is pertinent to note that the exclusion of these two nuclei causes  insignificant change in $\sigma$ from 1.645 MeV to 1.644 MeV. Moreover, exclusion of these two nuclei does not alter any of the parameter values listed in Table-1. However, the final calculations have been performed including all the nuclei.    

      In the Table-2 below, using the BWM mass formula with isotonic shifts beyond A=30, the atomic number Z and the neutron number N of the neutron and proton drip line nuclei have been tabulated along with the corresponding values of their one proton and one neutron separation energies (with errors in the parentheses). Uncertainty of 0.1 MeV in the calculation of binding energies have been used in the low mass region upto A=30 where isotonic shifts have not been incorporated. The new calculations predict $^{31}F$ and $^{8}He$ to be stable as can be seen in the Table-2. The values given in Table-2 could be useful for planning experiments on mass measurements. In order to demonstrate the change of sign of the one proton and one neutron separation energies, the atomic number Z and the neutron number N of nuclei with one neutron less than the proton drip line nuclei and nuclei with one neutron beyond the neutron drip line nuclei have also been listed along with the corresponding values of their one proton and one neutron separation energies. 

\begin{table}
\caption{Proton and neutron separation energies on and just beyond the proton and the neutron drip lines using the newly modified BW mass formula with isotonic shift.}
\begin{tabular}{cccccccccccc}
Proton& $S_p$&$S_n$ &One& $S_p$&$S_n$ &Neutron& $S_p$&$S_n$ &One& $S_p$&$S_n$   \\
&&&beyond&&&&&&beyond&& \\
&&&proton&&&&&&neutron&& \\ 
Drip    &         &           &drip &           &          &Drip     &          &          &drip&          &      \\ \hline
Z,N  &$ MeV$ & $MeV$&Z,N  &$ MeV$ & $MeV$ &Z,N  &$ MeV$ & $MeV$ &Z,N  &$ MeV$ & $MeV$  \\ \hline

  2,  1&   2.10( 11)&  26.70( 13)&  2,  0&  -8.20( 12)&  88.79( 32)&  2,  6&  31.10( 16)&   -.07( 14)&  2,  7&  33.81( 16)&  -3.18( 15)\\
  3,  2&   3.36( 12)&  19.61( 12)&  3,  1&  -3.49( 12)&  32.15( 15)&  3,  8&  26.96( 16)&    .58( 15)&  3,  9&  29.24( 17)&  -2.72( 16)\\
  4,  2&   1.11( 13)&  24.88( 14)&  4,  1&  -4.16( 12)&  35.92( 15)&  4, 10&  28.33( 17)&    .90( 17)&  4, 11&  30.38( 18)&  -2.50( 17)\\
  5,  3&    .74( 14)&  19.67( 14)&  5,  2&  -3.26( 13)&  28.90( 14)&  5, 12&  25.47( 18)&    .99( 17)&  5, 13&  27.26( 18)&  -2.47( 18)\\
  6,  3&    .17( 14)&  23.17( 15)&  6,  2&  -3.33( 15)&  32.29( 16)&  6, 14&  27.81( 19)&   1.01( 18)&  6, 15&  29.47( 19)&  -2.46( 18)\\
  7,  5&   1.98( 16)&  16.89( 16)&  7,  4&   -.90( 15)&  23.11( 15)&  7, 16&  25.08( 19)&    .97( 18)&  7, 17&  26.58( 19)&  -2.50( 18)\\
  8,  5&   1.98( 16)&  19.58( 16)&  8,  4&   -.71( 16)&  25.92( 16)&  8, 18&  27.74( 19)&    .94( 19)&  8, 19&  29.14( 19)&  -2.52( 19)\\
  9,  6&    .09( 16)&  21.03( 16)&  9,  5&  -2.33( 16)&  22.00( 17)&  9, 22&  28.26( 20)&    .33( 19)&  9, 23&  28.73( 20)&  -2.32( 19)\\
 10,  6&    .64( 17)&  23.33( 17)& 10,  5&  -1.67( 17)&  24.29( 17)& 10, 24&  28.16( 20)&    .31( 19)& 10, 25&  31.27( 20)&  -1.63( 19)\\
 11,  8&    .54( 17)&  20.00( 17)& 11,  7&  -1.49( 17)&  20.00( 17)& 11, 28&  29.27( 20)&    .03( 19)& 11, 29&  30.28( 20)&  -4.13( 19)\\
 12,  8&   1.37( 18)&  21.96( 18)& 12,  7&   -.59( 17)&  21.97( 18)& 12, 28&  28.13( 20)&    .64( 19)& 12, 29&  30.81( 20)&  -1.45( 19)\\
 13, 10&    .71( 18)&  19.39( 18)& 13,  9&  -1.05( 18)&  18.83( 18)& 13, 30&  26.84( 20)&    .83( 19)& 13, 31&  27.79( 20)&  -1.88( 19)\\
 14,  9&    .07( 18)&  20.56( 18)& 14,  8&  -1.66( 18)&  25.58( 18)& 14, 32&  29.45( 20)&    .53( 19)& 14, 33&  30.35( 20)&  -2.14( 19)\\
 15, 12&    .74( 18)&  18.98( 18)& 15, 11&   -.82( 18)&  18.05( 18)& 15, 32&  24.72( 20)&   1.65( 19)& 15, 33&  25.62( 20)&  -1.24( 19)\\
 16, 11&    .46( 18)&  19.60( 18)& 16, 10&  -1.08( 18)&  24.32( 18)& 16, 34&  27.39( 19)&   1.28( 19)& 16, 35&  28.24( 20)&  -2.71( 19)\\
 17, 14&   4.91( 18)&  22.92( 18)& 17, 13&   -.73( 19)&  17.51( 18)& 17, 34&  21.26( 19)&   2.02( 19)& 17, 35&  23.70( 19)&   -.27( 19)\\
 18, 13&   2.74( 19)&  20.95( 19)& 18, 12&   -.70( 19)&  23.41( 19)& 18, 46&  32.66( 20)&    .10( 18)& 18, 47&  33.35( 20)&  -2.97( 18)\\
 19, 16&    .55( 19)&  18.56( 18)& 19, 15&   -.97( 19)&  16.87( 19)& 19, 48&  29.14( 19)&    .01( 18)& 19, 49&  29.73( 19)&  -2.55( 18)\\
 20, 15&    .66( 19)&  17.90( 19)& 20, 14&   -.36( 19)&  23.60( 19)& 20, 50&  31.10( 19)&    .57( 18)& 20, 51&  31.51( 19)&  -4.29( 18)\\
 21, 18&    .44( 19)&  18.28( 18)& 21, 17&   -.95( 19)&  16.39( 19)& 21, 50&  27.84( 19)&   1.17( 18)& 21, 51&  27.68( 19)&  -4.44( 18)\\
 22, 17&    .64( 19)&  17.38( 19)& 22, 16&   -.35( 19)&  21.08( 19)& 22, 50&  28.93( 19)&   1.76( 18)& 22, 51&  29.52( 19)&  -3.85( 18)\\
 23, 20&    .04( 19)&  18.66( 18)& 23, 19&  -1.40( 19)&  16.46( 18)& 23, 50&  24.91( 19)&   2.37( 18)& 23, 51&  25.51( 19)&  -3.25( 18)\\
 24, 19&    .44( 19)&  17.33( 18)& 24, 18&   -.43( 19)&  20.34( 19)& 24, 50&  26.07( 18)&   2.96( 17)& 24, 51&  26.66( 18)&  -2.66( 18)\\
 25, 22&    .85( 19)&  18.95( 18)& 25, 21&  -1.26( 19)&  15.42( 18)& 25, 50&  22.16( 18)&   3.57( 17)& 25, 51&  22.77( 18)&  -2.06( 17)\\
 26, 20&    .30( 19)&  21.06( 18)& 26, 19&   -.34( 19)&  19.86( 19)& 26, 64&  36.37( 18)&    .12( 17)& 26, 65&  31.16( 18)&  -9.01( 17)\\
 27, 24&    .73( 18)&  18.79( 18)& 27, 23&   -.90( 18)&  16.21( 18)& 27, 64&  26.34( 18)&    .44( 17)& 27, 65&  32.76( 18)&  -2.59( 17)\\
 28, 21&    .24( 19)&  17.93( 18)& 28, 20&   -.91( 19)&  24.32( 18)& 28, 64&  27.76( 18)&    .96( 17)& 28, 65&  27.99( 18)&  -2.36( 17)\\
 29, 26&    .36( 18)&  18.76( 18)& 29, 25&  -1.03( 18)&  16.25( 18)& 29, 66&  25.15( 18)&    .00( 16)& 29, 67&  25.96( 18)&  -1.40( 16)\\
 30, 24&    .14( 18)&  20.58( 18)& 30, 23&   -.12( 18)&  19.27( 18)& 30, 70&  34.14( 18)&    .00( 16)& 30, 71&  28.87( 18)&  -8.22( 16)\\
 31, 29&    .55( 18)&  13.57( 18)& 31, 28&   -.42( 18)&  18.68( 18)& 31, 72&  31.89( 17)&    .32( 16)& 31, 73&  26.27( 17)&  -8.45( 16)\\
 32, 27&    .08( 18)&  17.12( 18)& 32, 26&   -.54( 18)&  20.41( 18)& 32, 72&  26.63( 17)&    .88( 16)& 32, 73&  33.56( 17)&  -1.52( 16)\\
 33, 32&   1.03( 18)&  15.20( 17)& 33, 31&   -.21( 18)&  13.57( 17)& 33, 74&  30.49( 17)&    .77( 16)& 33, 75&  32.11( 17)&   -.69( 16)\\
 34, 30&    .42( 18)&  17.76( 17)& 34, 29&   -.17( 18)&  15.92( 18)& 34, 78&  27.25( 17)&    .25( 16)& 34, 79&  27.63( 17)&  -2.12( 16)\\
 35, 34&    .88( 18)&  15.56( 17)& 35, 33&   -.41( 18)&  13.35( 17)& 35, 82&  25.51( 17)&    .02( 15)& 35, 83&  25.88( 17)&  -2.30( 15)\\
 36, 32&    .09( 18)&  17.64( 17)& 36, 31&   -.10( 18)&  15.48( 17)& 36, 84&  27.39( 17)&    .08( 15)& 36, 85&  27.75( 17)&  -2.21( 15)\\
 37, 36&    .75( 17)&  15.58( 17)& 37, 35&   -.18( 17)&  13.60( 17)& 37, 86&  24.97( 17)&    .15( 15)& 37, 87&  25.32( 17)&  -2.11( 15)\\
 38, 34&    .03( 17)&  18.10( 17)& 38, 33&   -.32( 18)&  15.59( 17)& 38, 88&  26.82( 16)&    .21( 15)& 38, 89&  27.15( 17)&  -2.03( 15)\\
 39, 38&    .77( 17)&  15.62( 17)& 39, 37&   -.22( 17)&  14.20( 17)& 39, 92&  25.11( 16)&    .01( 15)& 39, 93&  25.44( 16)&  -2.18( 15)\\
 40, 36&    .10( 17)&  18.07( 17)& 40, 35&   -.55( 17)&  15.83( 17)& 40, 94&  26.90( 16)&    .07( 15)& 40, 95&  27.22( 16)&  -2.10( 15)\\
 41, 40&    .89( 17)&  15.75( 16)& 41, 39&   -.37( 17)&  14.32( 16)& 41, 96&  24.59( 16)&    .13( 15)& 41, 97&  24.90( 16)&  -2.02( 14)\\
 42, 39&    .45( 17)&  14.94( 16)& 42, 38&   -.18( 17)&  18.41( 17)& 42, 98&  26.34( 16)&    .19( 14)& 42, 99&  26.65( 16)&  -1.94( 14)\\
 43, 42&    .66( 17)&  15.84( 16)& 43, 41&   -.40( 17)&  14.87( 16)& 43,102&  24.68( 16)&    .01( 14)& 43,103&  24.98( 16)&  -2.08( 14)\\
 44, 41&    .14( 17)&  15.47( 16)& 44, 40&   -.45( 17)&  18.62( 16)& 44,104&  26.38( 16)&    .07( 14)& 44,105&  26.67( 16)&  -2.00( 14)\\
 45, 44&    .34( 17)&  15.77( 16)& 45, 43&   -.67( 17)&  14.97( 16)& 45,106&  24.17( 16)&    .12( 14)& 45,107&  24.46( 16)&  -1.93( 14)\\
 46, 44&   1.78( 16)&  17.72( 16)& 46, 43&   -.17( 17)&  15.54( 16)& 46,108&  25.85( 16)&    .18( 14)& 46,109&  26.12( 16)&  -1.85( 14)\\
 47, 46&    .10( 16)&  15.75( 16)& 47, 45&   -.87( 16)&  15.38( 16)& 47,112&  24.22( 16)&    .02( 14)& 47,113&  24.50( 16)&  -1.98( 14)\\
 48, 46&   1.28( 16)&  17.50( 16)& 48, 45&   -.47( 16)&  15.93( 16)& 48,114&  25.85( 16)&    .07( 14)& 48,115&  26.12( 16)&  -1.91( 14)\\
 49, 49&    .18( 16)&  13.39( 15)& 49, 48&   -.05( 16)&  16.75( 16)& 49,116&  23.73( 15)&    .12( 14)& 49,117&  23.99( 15)&  -1.84( 14)\\
 50, 48&   1.04( 16)&  18.57( 15)& 50, 47&   -.78( 16)&  16.43( 16)& 50,118&  25.34( 15)&    .17( 13)& 50,119&  25.59( 15)&  -1.77( 13)\\
 51, 52&    .42( 16)&  13.77( 15)& 51, 51&   -.12( 16)&  11.75( 15)& 51,122&  23.76( 15)&    .03( 13)& 51,123&  24.01( 15)&  -1.89( 13)\\
 52, 52&   1.42( 16)&  15.29( 15)& 52, 51&   -.10( 16)&  12.24( 15)& 52,124&  25.32( 15)&    .07( 13)& 52,125&  25.56( 15)&  -1.82( 13)\\
 53, 57&    .10( 16)&  10.63( 15)& 53, 56&   -.16( 16)&  13.56( 15)& 53,126&  23.28( 15)&    .13( 13)& 53,127&  23.52( 15)&  -1.76( 13)\\
 54, 54&    .02( 16)&  14.46( 15)& 54, 53&   -.45( 16)&  12.88( 15)& 54,128&  24.82( 15)&    .17( 13)& 54,129&  25.06( 15)&  -1.69( 13)\\
 55, 61&    .17( 15)&  10.39( 15)& 55, 60&   -.11( 15)&  13.25( 15)& 55,132&  23.28( 15)&    .04( 13)& 55,133&  23.52( 15)&  -1.80( 13)\\
 56, 57&    .35( 16)&  12.08( 15)& 56, 56&   -.34( 16)&  14.89( 15)& 56,134&  24.79( 15)&    .08( 13)& 56,135&  25.01( 15)&  -1.74( 13)\\
 57, 63&    .25( 15)&  10.70( 14)& 57, 62&   -.09( 15)&  13.60( 14)& 57,136&  22.82( 15)&    .13( 13)& 57,137&  23.05( 15)&  -1.68( 13)\\
 58, 60&    .63( 15)&  15.05( 14)& 58, 59&   -.26( 15)&  12.09( 14)& 58,140&  24.74( 15)&    .01( 13)& 58,141&  24.95( 15)&  -1.78( 13)\\
 59, 65&    .05( 15)&  11.08( 14)& 59, 64&   -.24( 15)&  13.59( 14)& 59,142&  22.80( 15)&    .05( 13)& 59,143&  23.02( 15)&  -1.72( 13)\\
 60, 62&    .53( 15)&  15.43( 14)& 60, 61&   -.62( 15)&  13.29( 14)& 60,144&  24.26( 15)&    .10( 13)& 60,145&  24.47( 15)&  -1.66( 13)\\
 61, 68&    .14( 15)&  13.23( 14)& 61, 67&   -.29( 15)&  11.13( 14)& 61,146&  22.36( 15)&    .14( 13)& 61,147&  22.57( 15)&  -1.60( 12)\\
 62, 65&    .63( 15)&  12.92( 14)& 62, 64&   -.58( 15)&  15.59( 14)& 62,150&  24.20( 15)&    .03( 12)& 62,151&  24.40( 15)&  -1.70( 12)\\
 63, 71&    .02( 15)&  10.53( 14)& 63, 70&   -.15( 15)&  13.38( 14)& 63,152&  22.33( 15)&    .07( 12)& 63,153&  22.53( 15)&  -1.64( 12)\\
 64, 67&    .34( 15)&  13.09( 14)& 64, 66&   -.95( 15)&  15.66( 14)& 64,154&  23.73( 15)&    .11( 12)& 64,155&  23.93( 15)&  -1.59( 12)\\
 65, 74&    .04( 15)&  12.90( 14)& 65, 73&   -.40( 15)&  10.71( 14)& 65,158&  22.28( 15)&    .01( 12)& 65,159&  22.48( 15)&  -1.67( 12)\\
 66, 71&    .46( 15)&  12.21( 14)& 66, 70&   -.55( 15)&  15.00( 14)& 66,160&  23.66( 15)&    .05( 12)& 66,161&  23.85( 15)&  -1.62( 12)\\
 67, 78&    .36( 14)&  12.58( 13)& 67, 77&   -.09( 14)&  10.87( 13)& 67,162&  21.85( 15)&    .09( 12)& 67,163&  22.04( 15)&  -1.56( 12)\\
 68, 73&    .18( 15)&  12.48( 13)& 68, 72&   -.93( 15)&  15.05( 13)& 68,164&  23.21( 14)&    .13( 12)& 68,165&  23.40( 14)&  -1.51( 12)\\
 69, 80&    .11( 14)&  13.19( 13)& 69, 79&   -.28( 14)&  10.88( 13)& 69,168&  21.79( 14)&    .03( 12)& 69,169&  21.98( 14)&  -1.59( 12)\\
 70, 76&    .11( 14)&  14.79( 13)& 70, 75&   -.95( 14)&  13.09( 13)& 70,170&  23.13( 14)&    .07( 12)& 70,171&  23.31( 14)&  -1.54( 12)\\
 71, 84&    .33( 14)&  11.18( 13)& 71, 83&   -.07( 14)&   9.48( 13)& 71,172&  21.37( 14)&    .11( 12)& 71,173&  21.55( 14)&  -1.49( 12)\\
 72, 79&    .15( 14)&  12.75( 13)& 72, 78&   -.98( 14)&  15.12( 13)& 72,176&  23.04( 14)&    .02( 12)& 72,177&  23.21( 14)&  -1.57( 12)\\
 73, 86&    .09( 14)&  11.69( 13)& 73, 85&   -.31( 14)&   9.60( 13)& 73,178&  21.30( 14)&    .06( 12)& 73,179&  21.48( 14)&  -1.52( 12)\\
 74, 83&    .43( 14)&  10.32( 13)& 74, 82&   -.02( 14)&  14.77( 13)& 74,180&  22.60( 14)&    .10( 12)& 74,181&  22.77( 14)&  -1.47( 12)\\
 75, 90&    .20( 14)&  11.67( 13)& 75, 89&   -.15( 14)&   9.50( 13)& 75,184&  21.23( 14)&    .01( 12)& 75,185&  21.40( 14)&  -1.55( 12)\\
 76, 85&    .18( 14)&  11.24( 13)& 76, 84&  -1.09( 14)&  14.00( 13)& 76,186&  22.51( 14)&    .04( 12)& 76,187&  22.67( 14)&  -1.50( 12)\\
 77, 94&    .21( 14)&  11.54( 12)& 77, 93&   -.11( 14)&   9.29( 12)& 77,188&  20.82( 14)&    .08( 12)& 77,189&  20.99( 14)&  -1.45( 11)\\
 78, 88&    .07( 14)&  12.87( 12)& 78, 87&   -.24( 14)&  11.31( 13)& 78,190&  22.08( 14)&    .12( 11)& 78,191&  22.25( 14)&  -1.41( 11)\\
 79, 98&    .18( 14)&  11.44( 12)& 79, 97&   -.23( 14)&   9.06( 12)& 79,194&  20.74( 14)&    .04( 11)& 79,195&  20.90( 14)&  -1.47( 11)\\
 80, 92&    .03( 14)&  12.45( 12)& 80, 91&   -.12( 14)&  10.26( 12)& 80,196&  21.98( 14)&    .07( 11)& 80,197&  22.14( 14)&  -1.43( 11)\\
 81,102&    .25( 14)&  11.09( 12)& 81,101&   -.11( 14)&   9.09( 12)& 81,198&  20.34( 14)&    .11( 11)& 81,199&  20.50( 14)&  -1.38( 11)\\
 82, 96&    .17( 14)&  12.28( 12)& 82, 95&    .00( 14)&  10.91( 12)& 82,202&  21.88( 14)&    .03( 11)& 82,203&  22.03( 14)&  -1.45( 11)\\
 83,105&    .01( 14)&   8.81( 12)& 83,104&   -.05( 14)&  11.48( 12)& 83,204&  20.25( 14)&    .07( 11)& 83,205&  20.41( 14)&  -1.41( 11)\\
 84,101&    .27( 14)&   9.79( 12)& 84,100&   -.31( 14)&  12.41( 12)& 84,206&  21.47( 14)&    .10( 11)& 84,207&  21.62( 14)&  -1.36( 11)\\
 85,110&    .06( 13)&  11.03( 12)& 85,109&   -.17( 13)&   8.75( 12)& 85,210&  20.16( 14)&    .03( 11)& 85,211&  20.31( 14)&  -1.43( 11)\\
 86,105&    .10( 14)&   9.56( 12)& 86,104&   -.17( 14)&  12.41( 12)& 86,212&  21.36( 14)&    .06( 11)& 86,213&  21.50( 14)&  -1.38( 11)\\
 87,115&    .09( 13)&   8.58( 12)& 87,114&   -.04( 13)&  10.83( 12)& 87,214&  19.77( 14)&    .10( 11)& 87,215&  19.92( 14)&  -1.34( 11)\\
 88,109&    .21( 13)&   9.69( 12)& 88,108&   -.06( 13)&  11.84( 12)& 88,218&  21.24( 14)&    .02( 11)& 88,219&  21.38( 14)&  -1.40( 11)\\
 89,119&    .08( 13)&   8.47( 11)& 89,118&   -.02( 13)&  10.57( 12)& 89,220&  19.67( 14)&    .06( 11)& 89,221&  19.82( 14)&  -1.36( 11)\\
 90,112&    .04( 13)&  11.60( 12)& 90,111&   -.22( 13)&   9.72( 12)& 90,222&  20.84( 14)&    .09( 11)& 90,223&  20.98( 14)&  -1.32( 11)\\
 91,123&    .10( 13)&   8.40( 11)& 91,122&   -.04( 13)&  10.15( 11)& 91,226&  19.57( 14)&    .02( 11)& 91,227&  19.71( 14)&  -1.38( 11)\\
 92,116&    .12( 13)&  11.67( 11)& 92,115&   -.13( 13)&   9.71( 11)& 92,228&  20.72( 14)&    .05( 11)& 92,229&  20.86( 14)&  -1.34( 11)\\
 93,127&    .25( 13)&   6.94( 11)& 93,126&   -.03( 13)&   9.25( 11)& 93,230&  19.19( 14)&    .09( 11)& 93,231&  19.33( 14)&  -1.30( 11)\\
 94,120&    .18( 13)&  11.35( 11)& 94,119&   -.06( 13)&   9.34( 11)& 94,234&  20.60( 14)&    .02( 11)& 94,235&  20.73( 14)&  -1.35( 11)\\
 95,129&    .23( 13)&   7.17( 11)& 95,128&   -.01( 13)&   9.02( 11)& 95,236&  19.09( 14)&    .05( 11)& 95,237&  19.22( 14)&  -1.31( 11)\\
 96,124&    .23( 13)&  10.86( 11)& 96,123&   -.01( 13)&   9.54( 11)& 96,238&  20.21( 14)&    .09( 11)& 96,239&  20.34( 14)&  -1.27( 11)\\
 97,132&    .02( 13)&   8.98( 11)& 97,131&   -.21( 13)&   7.57( 11)& 97,242&  18.98( 14)&    .02( 11)& 97,243&  19.11( 14)&  -1.33( 11)\\
 98,127&    .04( 13)&   8.11( 11)& 98,126&   -.19( 13)&  10.40( 11)& 98,244&  20.09( 14)&    .05( 11)& 98,245&  20.22( 14)&  -1.29( 11)\\
 99,136&    .03( 13)&   9.58( 11)& 99,135&   -.19( 13)&   8.00( 11)& 99,246&  18.61( 14)&    .09( 11)& 99,247&  18.73( 14)&  -1.25( 11)\\
100,131&    .06( 13)&   8.25( 11)&100,130&   -.16( 13)&  10.19( 11)&100,250&  19.96( 14)&    .02( 11)&100,251&  20.09( 14)&  -1.30( 11)\\
101,140&    .04( 13)&   9.25( 11)&101,139&   -.18( 13)&   8.32( 11)&101,252&  18.49( 14)&    .06( 11)&101,253&  18.62( 14)&  -1.26( 11)\\
102,135&    .08( 13)&   8.66( 11)&102,134&   -.14( 13)&  10.31( 11)&102,254&  19.58( 14)&    .09( 11)&102,255&  19.71( 14)&  -1.23( 11)\\
103,131&    .45( 13)&   9.59( 11)&103,130&   -.41( 13)&  11.33( 11)&103,258&  18.38( 14)&    .03( 10)&103,259&  18.50( 14)&  -1.28( 10)\\
104,132&   2.57( 13)&  11.22( 11)&104,131&  -1.64( 13)&   9.81( 11)&104,260&  19.45( 14)&    .06( 10)&104,261&  19.58( 14)&  -1.24( 10)\\
105,134&   1.02( 13)&  11.11( 11)&105,133&  -3.18( 13)&   9.72( 11)&105,264&  18.26( 14)&    .00( 10)&105,265&  18.38( 14)&  -1.29( 10)\\
106,136&   2.66( 13)&  11.01( 11)&106,135&  -1.60( 13)&   9.62( 11)&106,266&  19.32( 14)&    .03( 10)&106,267&  19.44( 14)&  -1.25( 10)\\
107,137&    .52( 13)&   9.53( 11)&107,136&  -3.34( 13)&  11.23( 11)&107,268&  17.90( 14)&    .06( 10)&107,269&  18.02( 14)&  -1.22( 10)\\
108,138&   1.85( 13)&  11.13( 11)&108,137&  -2.00( 13)&   9.75( 11)&108,272&  19.19( 14)&    .01( 10)&108,273&  19.30( 14)&  -1.26( 10)\\
109,155&    .11( 13)&   7.63( 10)&109,154&   -.44( 13)&   8.65( 11)&109,274&  17.78( 14)&    .04( 10)&109,275&  17.89( 14)&  -1.23( 10)\\
110,141&    .97( 13)&   9.57( 11)&110,140&  -2.18( 13)&  11.24( 11)&110,276&  18.82( 14)&    .07( 10)&110,277&  18.94( 14)&  -1.19( 10)\\
111,160&    .21( 13)&   8.37( 10)&111,159&   -.13( 13)&   7.12( 10)&111,280&  17.65( 14)&    .02( 10)&111,281&  17.77( 14)&  -1.24( 10)\\
112,144&    .89( 13)&  11.04( 10)&112,143&  -2.36( 13)&   9.69( 10)&112,282&  18.69( 14)&    .04( 10)&112,283&  18.80( 14)&  -1.20( 10)\\
113,165&    .41( 13)&   6.70( 10)&113,164&   -.13( 13)&   7.90( 10)&113,284&  17.30( 14)&    .07( 10)&113,285&  17.41( 14)&  -1.17( 10)\\
114,148&    .95( 13)&  10.86( 10)&114,147&  -2.35( 13)&   9.52( 10)&114,288&  18.55( 14)&    .02( 10)&114,289&  18.66( 14)&  -1.21( 10)\\
115,170&    .18( 13)&   8.41( 10)&115,169&   -.01( 13)&   6.65( 10)&115,290&  17.18( 14)&    .05( 10)&115,291&  17.29( 14)&  -1.18( 10)\\
116,150&    .10( 13)&  10.97( 10)&116,149&  -2.73( 13)&   9.64( 10)&116,294&  18.41( 14)&    .00( 10)&116,295&  18.51( 14)&  -1.22( 10)\\
117,174&    .03( 13)&   8.45( 10)&117,173&   -.14( 13)&   6.95( 10)&117,296&  17.05( 14)&    .03( 10)&117,297&  17.16( 14)&  -1.19( 10)\\
118,154&    .23( 13)&  10.80( 10)&118,153&   -.27( 13)&   9.48( 10)&118,298&  18.05( 14)&    .06( 10)&118,299&  18.16( 14)&  -1.16( 10)\\

\end{tabular} 
\end{table}

\begin{figure}[h]
\eject\centerline{\epsfig{file=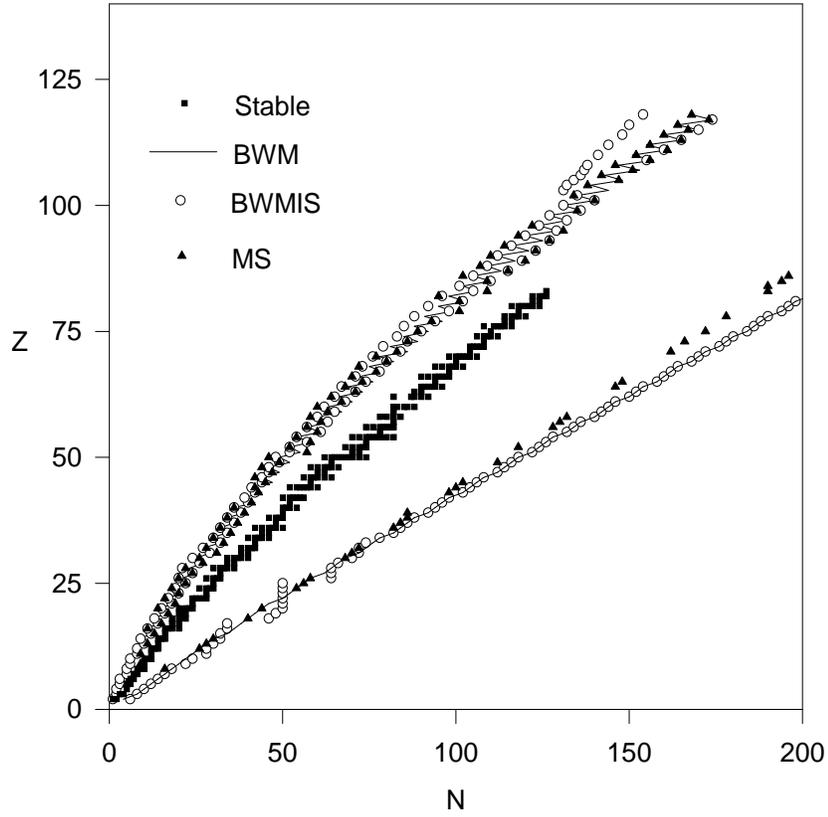,height=20cm,width=15cm}}
\caption
{The atomic number Z has been plotted against the neutron number N as solid squares for the stable nuclei and the driplines have been calculated using the BWM mass formula. The solid triangles represent dripline nuclei using MS masses while the hollow circles represent dripline nuclei using BWM mass formula with the isotonic shifts beyond A=30.}
\label{fig1}
\end{figure}

\begin{figure}[h]
\eject\centerline{\epsfig{file=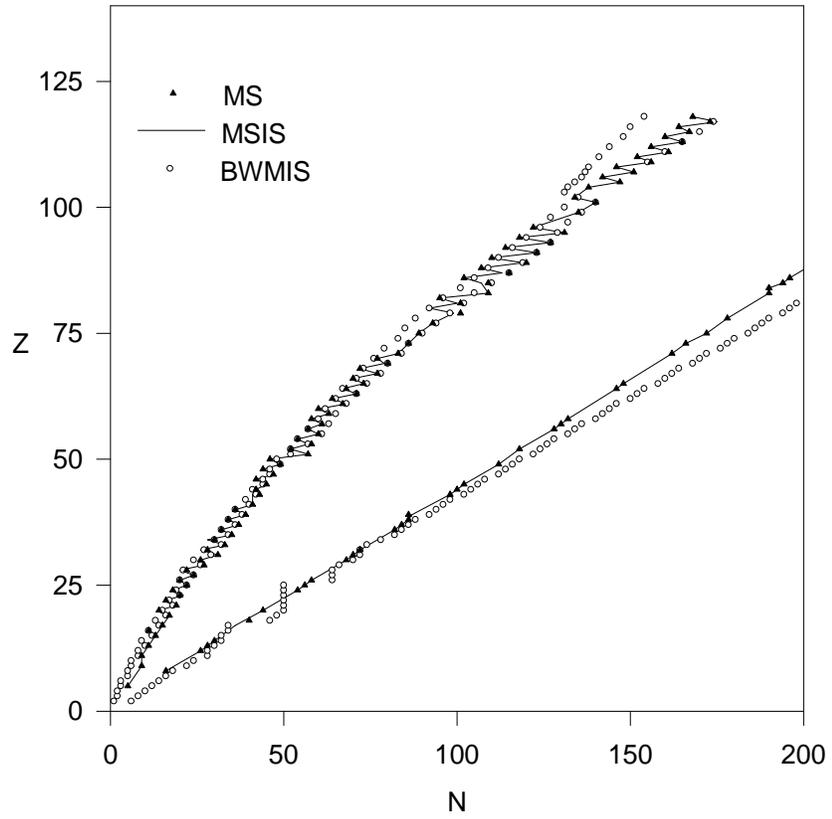,height=20cm,width=15cm}}
\caption
{The driplines have been calculated using the MS masses with the isotonic shifts. The solid triangles represent dripline nuclei using MS masses while the hollow circles represent dripline nuclei using BWM mass formula with the isotonic shifts beyond A=30.}
\label{fig2}
\end{figure}

      In Fig.~\ref{fig1}, using the BWM mass formula, the atomic number Z has been plotted against the neutron number N for the proton and neutron dripline nuclei. The lines joining the proton dripline nuclei and the neutron dripline nuclei have been shown as the proton dripline and the neutron dripline respectively in the figure. The nuclei stable against decay via $\beta^{\pm}$, $\alpha$ or heavy particle emissions or electron capture or spontaneous fission \cite{r10} have been shown with small solid squares. There is no nucleus beyond the atomic number Z=83 which is stable against the decays via $\beta^{\pm}$, $\alpha$, heavy particle emissions and electron capture and spontaneous fission. The hollow circles depict the proton and the neutron dripline nuclei using the BWM mass formula with isotonic mass shift corrections (BWMIS) beyond A=30 while the solid triangles represent dripline nuclei using MS masses. In Fig.~\ref{fig2}, the driplines have been calculated using the MS masses with the isotonic shifts (MSIS). The solid triangles represent dripline nuclei using MS masses while the hollow circles represent dripline nuclei using BWM mass formula with the isotonic shifts beyond A=30. 
Fig.~\ref{fig2} shows that both the driplines with MS and MSIS are almost the same except at few scattered points. Some of the MS masses near both the driplines are not available. Specifically, MS-predictions for some even Z proton dripline nuclei are not available around N=90 and Z=75. As there is no simple formula by which the 
missing masses of MS can be calculated easily, they are missing in the figures. Interestingly, the driplines predicted by BWMIS are almost similar to those obtained by MSIS except at large N or Z values where we find that BWMIS gives better mass predictions, that is, closer to available experimental masses than MS. 
  
      In summary, the atomic mass excesses have been calculated using the modified Bethe-Weizs\"acker mass formula with isotonic mass shifts incorporated and compared with the predictions of the mass formula of Myers-Swiatecki. It is found that the agreements between the experimental masses and the calculated ones 
are greatly improved by introducing the isotonic mass shifts. The  $\chi^2/N$= 0.907 predicted by BWMIS is comparable to the $\chi^2/N$= 1.398 predicted by MS. Neutron and proton separation energies have been calculated using the modified Bethe-Weizs\"acker mass formula with isotonic mass shifts. The  last bound isotopes of neutron and proton rich nuclei have been identified and the neutron and proton driplines have been provided. The significant changes by incorporating isotonic mass shifts have been observed for the proton dripline near heavier mass region. It is pertinent to note that BWM is a simple mass formula compared to the mass formula of Myers and Swiatecki. Nevertheless, with isotonic shift incorporated the modified Bethe-Weizs\"acker mass formula not only predicts almost the same driplines as the MS formula but provides better mass predictions for some nuclei with higher N, Z values as well. These observations are important as they suggest that the simple BWMIS prescription would be useful for theoretical computations of masses which are used in various computational codes.

\end{document}